\documentclass{aa}
\usepackage{graphics}

\begin{document}

   \thesaurus{03         
      (11.09.1 NGC~4522; 
       11.09.2;  
       11.09.4;  
       11.11.1)  
              }

   \title{The consequences of ram pressure stripping on the Virgo cluster
spiral galaxy NGC~4522}

   \author{B.~Vollmer\inst{1,3}, M.~Marcelin\inst{2}, P.~Amram\inst{2}, C.~Balkowski\inst{1}, V.~Cayatte\inst{1} \and O.~Garrido\inst{2}}

   \offprints{B.~Vollmer, e-mail: Bernd.Vollmer@obspm.fr}

   \institute{Observatoire de Paris, DAEC,
              UMR 8631, CNRS et Universit\'e Paris 7,
      F-92195 Meudon Cedex, France. \and
      Observatoire de Marseille, 2 Place le Verrier,
      F-13248 Marseille Cedex 04, France. \and
      Institut f\"ur Theoretische
              Astrophysik der Universit\"at Heidelberg,
      Tiergartenstra{\ss}e 15, D-69121 Heidelberg, Germany.
      }

   \date{Received / Accepted}

   \authorrunning{Vollmer et al.}
   \titlerunning{The consequences of ram pressure stripping on NGC~4522}

   \maketitle

\begin{abstract}
New H$\alpha$ Fabry-Perot interferometer observations of the Virgo cluster
spiral galaxy NGC~4522 are presented. A velocity field up to a radius of
60$''$ was obtained. The observed rotation curve is symmetric.
It is not perturbed in the inner disk. For radii $>50''$ it shows a solid body
rotation. We compare these data with a numerical
model which includes the effects of ram pressure stripping.
The model can reproduce the main characteristics
of the H$\alpha$ emission distribution and velocity field. We therefore
conclude that the stripping event which caused the H{\sc i} deficiency
and the distorted H$\alpha$ distribution and velocity field has been due to
ram pressure. The closest passage of the galaxy to the cluster center
is estimated to be $\sim$6.5\,10$^{8}$ yr ago. The observed recent star
formation is due to the re-accretion of gas clouds which were accelerated
to velocities below the escape velocity during the stripping event.
Furthermore, the gas which is located at radii $>60''$ is part of an
asymmetric expanding ring of gas clouds.
A substantial part of these clouds is located above the disk
plane. We predict the H{\sc i} gas distribution and velocity field as it would
be observed by deep 21 cm observations.

\end{abstract}


\section{Introduction}

It is a well established fact that spiral galaxies in clusters have less
atomic gas than isolated spirals of the same morphological type and same
optical diameter, i.e. they are H{\sc i} deficient (Chamaraux et al. 1980,
Bothun et al. 1982, Giovanelli \& Haynes 1985, Gavazzi 1987, 1989).
There are mainly two kinds of mechanisms which are able to cause the removal
of the atomic gas: (i) tidal interactions or (ii) the interaction of the
interstellar medium
(ISM) with the hot intracluster medium (ICM). The mapping of the gas content
of spiral galaxies in the Virgo (Cayatte et al. 1990, 1994) and Coma cluster
(Bravo-Alfaro et al. 2000) showed that the H{\sc i} disk sizes of cluster
spirals are considerably reduced. In addition, galaxies with a symmetric
optical disk that have an asymmetric H{\sc i} gas distribution are quite
frequent in the cluster core. These observational results indicate that
the gas removal due to the rapid motion of the galaxy within the ICM
(ram pressure stripping; Gunn \& Gott 1972) is responsible for the H{\sc i}
deficiency and the distorted gas disks of the cluster spirals. Nevertheless,
it has not yet been unambiguously shown that ram pressure stripping
is responsible for the H{\sc i} distribution and kinematics of one given
spiral galaxy. The possibility of a tidal interaction producing the
distortion has never been completely ruled out.

Very few simulations
have been done to quantify ram pressure stripping (Gaetz et al. 1987,
Balsara et al. 1994, Tosa 1994, Abadi et al. 1999). All of them have
considered that the Interstellar medium (ISM) is continuous.
In order to take into account the clumpiness of the ISM, we use a sticky
particle model in which each particle represents a cloud complex with an
assigned mass dependent radius. The viscosity of the clumpy ISM is due
to inelastic collisions between the particles. The effect of ram pressure is
modeled as an additional acceleration applied on the particles located at the
front side of the galaxy motion. The gas distribution
and the velocity field of a given simulation snapshot can be directly compared
with observations.

In the Virgo cluster only few spiral galaxies with peculiar gas distributions
were studied in detail:
\begin{itemize}
\item
NGC~4419: Kenney et al. (1990) observed an asymmetric CO distribution in this
cluster spiral. They suggest that its ISM is undergoing a strong interaction
with the ICM.
\item
NGC~4254: Phookun \& Mundy (1993) observed a counter-rotating, very
low surface density H{\sc i} plume.
\item
NGC~4654: Phookun \& Mundy (1995) found a very asymmetric H{\sc i}
distribution, with a
compressed edge on one side and a long tenuous tail on the other.
They conclude that this is a result of the combination of rotation and
ram pressure.
\item
NGC~4388: Veilleux et al. (1999) made H$\alpha$ and [O{\sc iii}]
$\lambda$5007 observations with the Hawaii Imaging Fabry-Perot Interferometer
(HIFI). They found a large complex of highly ionized gas that extends well
above the disk. A ram pressure model with a Mach cone of opening angle
$\sim$80$^{\rm o}$ is proposed to explain the observed distribution and
kinematics of the ionized gas.
\end{itemize}

The H{\sc i} deficient spiral galaxy NGC~4522 was observed in the optical
and H$\alpha$ with the WIYN telescope.
Kenney \& Koopmann (1999) showed that its old stellar disk is relatively
undisturbed. However, the H$\alpha$ distribution is very peculiar.
Filaments emerge from the outer edge of the disk which is abruptly truncated
beyond 0.35$R_{25}$. They suggest that this features are due to a bow shock
caused by ram pressure which is still acting on the galaxy. Nevertheless,
they made their conclusion only on morphological grounds.

With the imaging of the gas content of galaxies only two ($\alpha$, $\delta$)
out of six components ($x$, $y$, $z$, $v_{\rm x}$, $v_{\rm y}$, $v_{\rm z}$)
of phase space are accessible. The knowledge of the velocity field
represents a considerable increase of information. But still, the
derived three-dimensional distribution and velocity field depends
strongly on the applied model. The number
of possible model solutions decreases strongly with the number of observed
quantities. Thus, the knowledge of both the gas distribution and
velocity field gives very strong constraints on dynamical models.

We therefore observed NGC~4522 with the Fabry-Perot Interferometer
at the Observatoire de Haute Provence (OHP). We were able to recover
its velocity field up to a radius of $\sim$60$''$.
We first present the observations (Section~2) and the numerical model
(Section~3). The observational results are compared to the model
in Section~4. We discuss the outcomes of the comparison and draw
our conclusions in Section~5.

We adopt a distance of 17 Mpc for the Virgo cluster.

\section{Observations and data reductions}

\subsection{The GHASP Survey}

The Gassendi HAlpha survey of SPiral galaxies
\footnote{http://www-obs.cnrs-mrs.fr/interferometrie/ghasp.html}
consists of mapping the
H$\alpha$ distribution and kinematics of a sample of about 200
nearby spirals using a scanning Fabry Perot interferometer (FP).
The aim of the survey begun in October 1998
is to obtain the kinematics with both high spatial and spectral resolutions
in the optical disk for a homogeneous sample of galaxies.
This should allow us to tune the parameters of the mass
distribution accurately and to study the internal dynamics of disks.
This project is complementary to the Westerbork Survey
of H{\sc i} in Spiral Galaxies\footnote{http://thales.astro.rug.nl/~whisp/}.

The FP is placed in a focal reducer
bringing the original f/15 focal ratio of the Cassegrain focus to f/3.8.
The focal reducer is attached at the Cassegrain
focus of the 1.93 m OHP telescope (Observatoire de Haute
Provence, France).
Narrow band and high transmission interference filters are
used to isolate the studied emission line.
The detector used is an image photon counting system (IPCS).
The IPCS, with a time sampling of 1/50 s and zero readout
noise makes it possible to scan the interferometer rapidly (typically 10
s per channel) avoiding sky transparency, airmass and seeing variation
problems during the exposures.
The field of view is 4.1 arcmin x 4.1 arcmin and the pixel size 0.96 arcsec.
A full description of the GHASP survey and equipment
will be given in a forthcoming paper but the basic principles of a
similar instrument can be found in Amram et al. (1991).

\subsection{Observations}

The observation of NGC~4522 have been made with the GHASP equipment
in April 1999.
Using a scanning FP of interference order 796 at H$\alpha$ the free
spectral range of 378 km\,s$^{-1}$ was scanned through 24 channels
(the finesse of
the interferometer being 12) providing a spectral sampling of 17
km\,s$^{-1}$.
The journal of the observations is given in Table~\ref{tab:journal}.

\begin{table}
      \caption{Journal of Perot-Fabry observations}
         \label{tab:journal}
      \[
         \begin{array}{ll}
            \hline
            \noalign{\smallskip}
            {\rm Observations} & \\
    \hline
            \noalign{\smallskip}
       {\rm Telescope} & {\rm OHP\ 1.93\ m} \\
       {\rm Equipment} & {\rm GHASP\ @\ Cassegrain\ focus} \\
       {\rm Date} & {\rm April,\ 7\ and\ 13\ 1999} \\
               {\rm Seeing} & $$\sim$$ 2" \\
            \noalign{\smallskip}
            \hline
            \noalign{\smallskip}
            {\rm Interference\ Filter}  & \\
    \hline
            \noalign{\smallskip}
    {\rm Central\ Wavelength} & 6615 {\rm \AA} \\
                    {\rm FWHM} & 11 {\rm \AA} \\
                    {\rm Transmission} & 0.6 \\
            \noalign{\smallskip}
            \hline
    \noalign{\smallskip}
    {\rm Calibration} &  \\
    \hline
            \noalign{\smallskip}
  {\rm Neon\ Comparison\ light} & $$\lambda$$ 6598.95 {\rm \AA} \\
    \noalign{\smallskip}
            \hline
    \noalign{\smallskip}
    {\rm Fabry-Perot} &  \\
    \hline
            \noalign{\smallskip}
  {\rm Interference\ Order} & 796 @ 6562.78 {\rm \AA} \\
                  {\rm Free\ Spectral\ Range\ at\ H}$$\alpha$$ & 380\ {\rm km\,s}$$^{-1}$$ \\
                  {\rm Finesse\ at\ H}$$\alpha$$ & 12 \\
                  {\rm Spectral\ resolution\ at\ H}$$\alpha$$ & 18750\ {\rm at\ the\ sample\ step} \\
    \noalign{\smallskip}
            \hline
    \noalign{\smallskip}
    {\rm Sampling} &  \\
     \hline
            \noalign{\smallskip}
 {\rm Number\ of\ Scanning\ Steps} & 24 \\
          {\rm Sampling\ Step} & 0.35 {\rm \AA}\ (16\ {\rm km\,s}$$^{-1}$$) \\
          {\rm Total\ Field} & 4.1'$$\times $$4.1' (256$$\times $$256 px$$^{2}$$) \\
          {\rm Pixel\ Size} & 0.96''  \\
    \noalign{\smallskip}
            \hline
    \noalign{\smallskip}
    {\rm Exposures\ times}  & \\
    \hline
            \noalign{\smallskip}
 {\rm Total\ exposure} & 3.5\ {\rm hours} \\
                 {\rm Elementary\ scanning} & \\
 {\rm exposure\ time} & {\rm 10\ s\ per\ channel} \\
                 {\rm Total\ exposure\ time} & \\
 {\rm per\ channel} & 525\ {\rm s}\\
    \noalign{\smallskip}
            \hline
         \end{array}
      \]
\end{table}

\begin{table}
      \caption{Physical Parameters of NGC~4522}
         \label{tab:parameters}
      \[
         \begin{array}{ll}
            \hline
            \noalign{\smallskip}
{\rm Other\ names} &  {\rm UGC~7711} \\
& {\rm VCC~1516} \\
& {\rm CGCG~070-168} \\
$$\alpha$$\ (1950)$$^{\rm a}$$ &  12$$^{\rm h}31^{\rm m}7.62^{\rm s}$$\\
$$\delta$$\ (1950)$$^{\rm a}$$ &  9$$^{\rm o}27'3.1''$$\\
{\rm Morphological\ type}$$^{\rm a,\ b}$$ & {\rm SBcd,\ Sc/Sb} \\
{\rm Optical\ diameter\ D}_{25}$$^{\rm a}$$\ ($$'$$) & 3.7\\
{\rm B}$$_{T}^{0}$$$$^{\rm a}$$ & 12.02\\
{\rm Systemic\ heliocentric\ velocity}$$^{\rm c}$$\ {\rm (km\,s}$$^{-1}$$) & 2337$$\pm$$5\\
{\rm (optical)} & \\
{\rm HI\ heliocentric\ velocity}$$^{\rm d}$$\ {\rm (km\,s}$$^{-1}$$) & 2330$$\pm$$5 \\
{\rm Distance\ D\ (Mpc)} & 17 \\
{\rm Log\ L(H}$$\alpha$$)$$^{\rm e}$$\ ({\rm erg\,s}$$ ^{-1}$$) & 40.11\\
{\rm Vrot}$$_{\rm max}^{\rm f}$$\ {\rm (km\,s}$$^{-1}$$) & 110$$\pm$$10 \\
{\rm R}$$_{\rm max}^{\rm f}$$\ {\rm (kpc)} & 4.5$$\pm$$0,5 \\
{\rm M}$$_{\rm tot}^{\rm g}$$\ ($$\times$$ 10$$^{10}$$\ {\rm M}$$_{\odot}$$) & 2 \\
{\rm PA\ (gas\ kinematics)}$$^{\rm f}$$ & 33$$^{\rm o}$$ \\
{\rm PA\ (optical\ image)}$$^{\rm a}$$ & 33$$^{\rm o}$$ \\
{\rm Inclination\ of\ the\ gaseous\ disk}$$^{\rm e}$$ & 75$$^{\rm o} \pm 5^{\rm o}$$ \\
{\rm FWHM\ of\ central\ profiles}$$^{\rm f}$$\ {\rm (km\,s}$$^{-1}$$) &80$$\pm$$10 \\
{\rm HI\ deficiency}$$^{\rm h}$$ & 0.51 \\
\noalign{\smallskip}
\hline
\end{array}
      \]
\begin{list}{}{}
\item[$^{\rm{a}}$] RC3
\item[$^{\rm{b}}$] Binggeli et al. (1985)
\item[$^{\rm{c}}$] Rubin et al. (1998)
\item[$^{\rm{d}}$] Helou et al. (1984)
\item[$^{\rm{e}}$] Kenney \& Koopmann (1999)
\item[$^{\rm{f}}$] this paper
\item[$^{\rm{g}}$] inside D$_{25}$ assuming a flat rotation curve of $V_{\rm max}$=100 km\,s$^{-1}$
\item[$^{\rm{h}}$] Giovanelli \& Haynes (1985)
\end{list}
\end{table}

\subsection{Reductions}

The data were reduced using the ADHOCw software package
\footnote{ftp availaible
http://www-obs.cnrs-mrs.fr/adhoc/adhoc.html, Boulesteix et al. (1999)}.
The data reduction procedure has been
extensively described in Amram et al. (1996) and references
therein.

Wavelength calibrations were obtained by scanning the narrow Ne 6599
\AA\ line under the same conditions as the observations.  The relative
velocities with respect to the systemic velocity are very accurate,
with an error of a fraction of a channel width ($<$3 km\,s$^{-1}$)
over the whole field.

The signal measured along the scanning sequence was separated into
two parts: (1) an almost constant level produced by the continuum light
in a 10 \AA ~passband around H$\alpha $ (continuum map), and (2) a
varying part produced by the H$\alpha $ line (monochromatic map).  The
continuum level was taken to be the mean of the three faintest
channels, to avoid channel noise effects.  The monochromatic map was
obtained by integrating the monochromatic profile in each pixel.  The
velocity sampling was 16 km\,s$^{-1}$.  The monochromatic maps had
one-pixel resolution in the center of the galaxy. Spectral profiles
were binned in the outer parts (to 5 $\times$ 5 pixels) in order to
increase the signal-to-noise ratio.  When multiple components were
visually present, the lines were decomposed into multiple Gaussian
components.  OH night sky lines passing through the filters were
subtracted by determining the shapes and intensities of the lines away
from the galaxies (Laval et al. 1987).

\section{The numerical model}

\begin{figure}
\resizebox{\hsize}{!}{\includegraphics{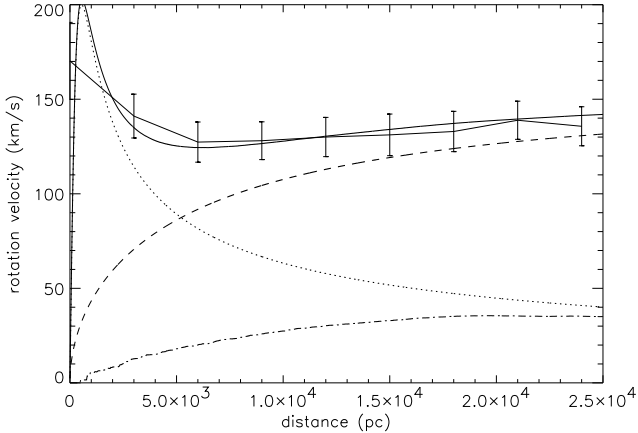}}
\caption{Initial rotation curve for the simulation.
Dashed: halo, dotted: bulge, dash-dotted: gas,
thick: total, solid: initial model rotation curve.
The error bars correspond to the model cloud velocity dispersion.
} \label{fig:initrotationvel}
\end{figure}

We used the three-dimensional N-body code described in detail in Vollmer
et al. (2000). The particles represent gas cloud complexes which are
evolving in an analytically given gravitational potential of the galaxy.
This potential consists of two spherical parts: the dark matter halo
and the stellar bulge. The outcoming velocity field has a constant rotation
curve of $v_{\rm rot} \sim$130 km\,s$^{-1}$ (Fig.~\ref{fig:initrotationvel}).
10\,000 particles of different masses are rotating within this gravitational
potential. A radius, which depends on the mass, is attributed to each
particle. During the disk evolution they can have inelastic collisions.
The outcome of these collisions is simplified following Wiegel (1994):
\begin{itemize}
\item
for $r_{1}-r_{2} < b < r_{1}+r_{2}$:\\ fragmentation
\item
for $b \le r_{1}-r_{2}$ and $v_{\rm esc} > v_{\rm f}$:\\
mass exchange
\item
for $b \le r_{1}-r_{2}$ and $v_{\rm esc} \le v_{\rm f}$:
\\ coalescence,
\end{itemize}
where $b$ is the impact parameter, $r_{1}$ and $r_{2}$ the cloud radii,
$v_{\rm f}$ the final velocity difference, and $v_{\rm esc}$ is the escape
velocity. This results in an effective gas viscosity in the disk.

As the galaxy moves through the ICM its clouds are accelerated by
ram pressure. Within the galaxy's inertial system the clouds
are exposed to a wind coming from the front side of its motion.
The effect of ram pressure on the clouds is simulated by an additional
force on the clouds in the wind direction. Only clouds which
are not protected by other clouds against the wind are affected.

As the galaxy approaches the cluster center its velocity increases.
At the same time the surrounding ICM density increases. This leads
to an increase of the ram pressure on the ISM clouds $p_{\rm ram}=
\rho_{\rm ICM} v_{\rm gal}^{2}$, where $\rho_{\rm ICM}$ is the ICM
density and $v_{\rm gal}$ is the velocity of the galaxy.
We take this evolution of $p_{\rm ram}$ into account in adopting
the following profile $p_{\rm ram}=28\,\rho_{0} v_{0}^{2} {\rm exp} \big(
-(t/8\,10^{7})^{2} \big)$, where $\rho_{0}=10^{-4}$ cm$^{-3}$ and
$v_{0}$=1000 km\,s$^{-1}$. This corresponds to a galaxy's closest approach
of $\sim 10^{5}$ pc to the cluster center.

The choice of the maximal ram pressure and the inclination angle $i$
between the disk and the orbital plane
depends on the resulting H{\sc i} deficiency and the fraction of re-accreted
gas (see Section \ref{sec:results}).
Furthermore, the observed gas distribution gives an indication
for the right range of the inclination angle $i$.
The chosen parameters $i$=20$^{\rm o}$
and $p_{\rm ram}^{\rm max}=28\,\rho_{0} v_{0}^{2}$ lead to a model H{\sc i}
deficiency of $DEF$=0.4 compared to the observed deficiency of $DEF$=0.5.
The correspondence is good enough in the way that the simulation
shows all characteristics of the distribution and kinematics necessary
to identify the stripping process unambiguously.
We have let the galaxy evolve during 10$^{9}$ yr before beginning the ram
pressure simulations.

\section{Results\label{sec:results}}

\begin{figure}
\resizebox{10cm}{!}{\includegraphics{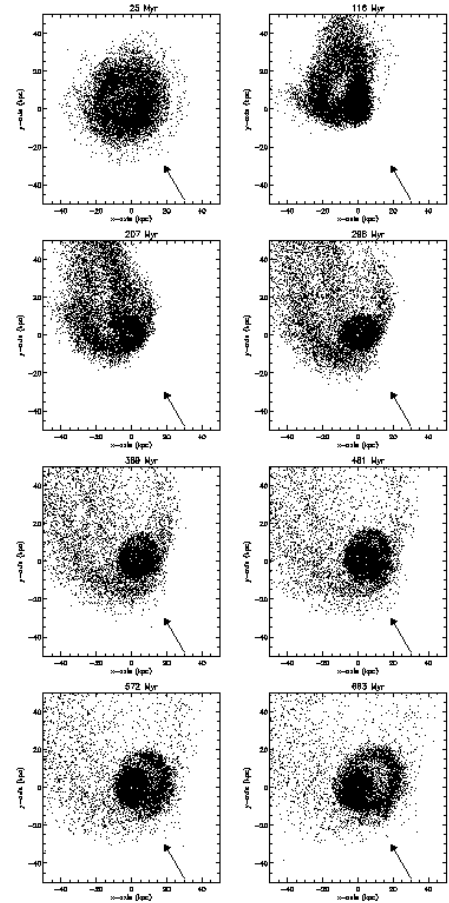}}
\caption{Snapshots of the simulation. The elapsed time is indicated
at the top of each panel.
The galaxy is seen face-on and is moving in the south-west
direction, i.e. the wind is coming from the south-west. This wind
direction is indicated by the arrows. The galaxy rotates
counter-clockwise.
} \label{fig:film}
\end{figure}

A sequence of the galaxy evolution is shown in Fig.~\ref{fig:film}.
At $t_{0}=6\,10^{8}$ yr ram pressure is at maximum. An overdensity builds
up within the disk in the wind direction (south-west) together with
an asymmetric ring structure.
About 10$^{8}$ yr later, the material in the ring moves to the north-east
leaving the disk. At this stage of the evolution ram pressure has
already ceased and the subsequent evolution of the ISM is mainly governed by
rotation and re-accretion. As for the case of edge-on stripping, there
is an accelerated arm where the wind pushes the clouds in the sense
of rotation and a decelerated arm where the clouds are decelerated by the wind
(see Vollmer et al. 2000).
The accelerated arm extends to the north, the decelerated arm is horizontal.
After $t-t_{0} \sim 2\,10^{8}$ yr the clouds, which where not
accelerated to the escape velocity, begin to fall back onto the galaxy.
This happens within two distinct arms. A small, counter-rotating arm
in the north-west and a more massive arm coming from the
north-east where the clouds fall back to the galaxy in the sense of rotation.
It hits the outer galactic disk in the south-west carrying a part of
the disk gas clouds with it to larger galactic radii. This leads to the
formation of an expanding asymmetric ring at $t-t_{0} \sim 3\,10^{8}$ yr.
During the expansion the re-accreting gas clouds of the counter-rotating arm
collide with the clouds located in the ring. We suggest that these collisions
lead to star formation activity within the ring structure.

Vollmer et al. (2000) have shown that in general a substantial part of the
stripped ICM re-accretes between 2 and 7\,10$^{8}$ yr after the closest passage
of the galaxy to the cluster center. Furthermore, the effects of ram pressure
within a Virgo spiral galaxy can only be observed after its passage in the
cluster core. This is also the case for the simulation shown
in Fig.~\ref{fig:film}. The large projected distance of
NGC~4522 with respect to the Virgo cluster center (M87) of 3.3$^{\rm o}$
($d \sim$1 Mpc) makes it thus very unlikely that the galaxy is just entering
the cluster. The ICM density at this distance is only
$\rho_{\rm ICM}\sim 3\,10^{-5}$ cm$^{-3}$ (Schindler et al. 1999).
Even if the galaxy has a velocity of 1800 km\,s$^{-1}$ the resulting
ram pressure does not exceed $\rho_{0} v_{0}^{2}$. For these low values
the influence of ram pressure on the neutral component of the ISM is not
observable.

We therefore come to the conclusion that NGC~4522 has already
passed the cluster center.
Since the galaxy is leaving the cluster now, we see the consequences of
a past strong stripping event. Assuming a mean galaxy velocity of
$v_{\rm gal}$=1500 km\,s$^{-1}$ we can calculate the time elapsed
since its closest passage to the cluster center:
$t\simeq d/v_{\rm gal}\simeq$ 6.5\,10$^{8}$ yr. We therefore compare a
snapshot of our simulation at this timestep with our observations.
In this stage most of the re-accretion has already taken place.
Since our model does not explicitly include star formation we can only
speculate that this re-accretion has lead to
an increase of the star formation rate making the galaxy bluer.
At the present day, accretion is still happening and the collisions between the
infalling clouds and the ISM probably still induces star formation.
For direct comparison we use the position angle, inclination and systemic
velocity of NGC~4522 (Table~\ref{tab:parameters}).

\begin{figure}
\resizebox{8.8cm}{!}{\includegraphics{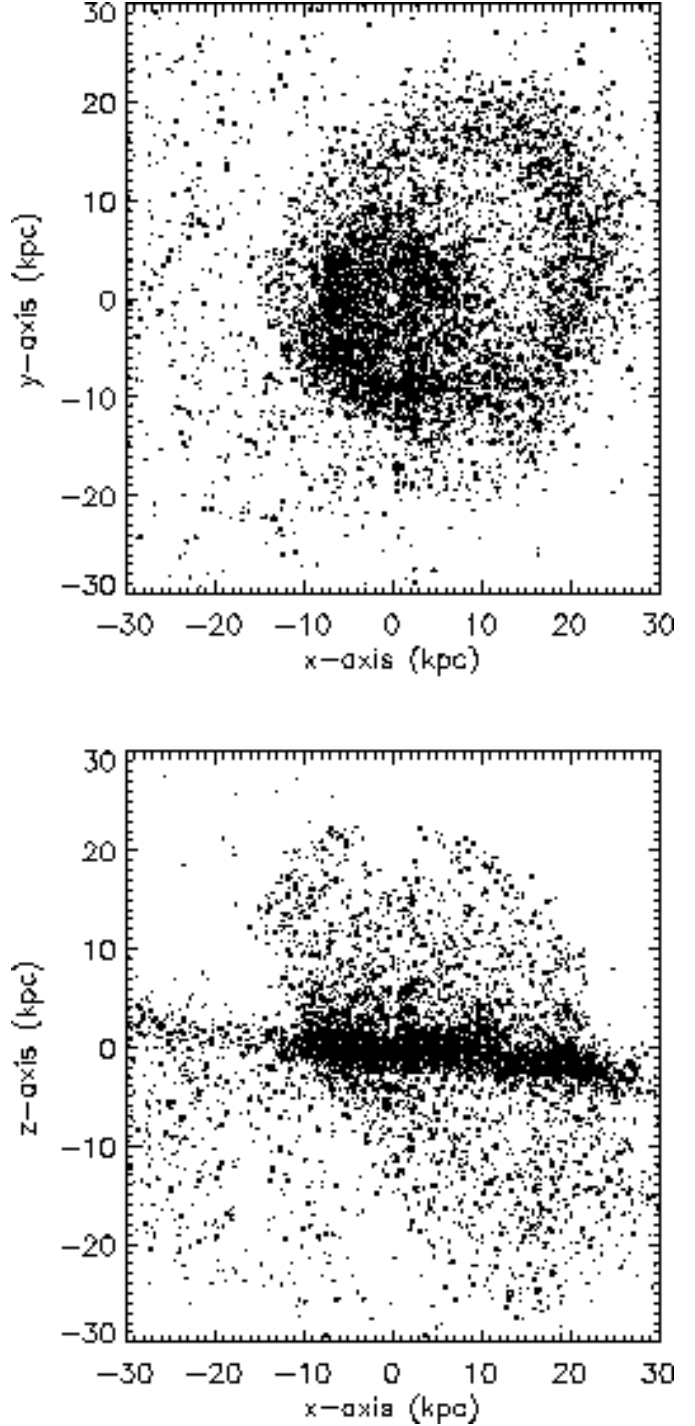}}
\caption{The gas cloud distribution of the galaxy. The radii of the
circles correspond to the cloud radii. Upper panel: Face-on view.
Lower panel: Edge-on view.
} \label{fig:point_dist}
\end{figure}

First, we show in Fig.~\ref{fig:point_dist} the three-dimensional particles
distribution and velocity
field of the snapshot of our simulation that we want to compare with
the observations. It shows the gas cloud distribution
of the galaxy in a face-on and edge-on view. It corresponds to the
last snapshot of Fig.~\ref{fig:film}.
The H{\sc i} disk has a diameter of $\sim$14 kpc. The asymmetric ring
extends up to radii of $\sim$20 kpc. A part of the ring is located
beyond the disk plane (Fig.~\ref{fig:point_dist} lower panel).
At this stage of evolution, a tenuous gas component
forms complex three-dimensional structure around the galaxy disk.

\begin{figure}
\resizebox{8.8cm}{!}{\includegraphics{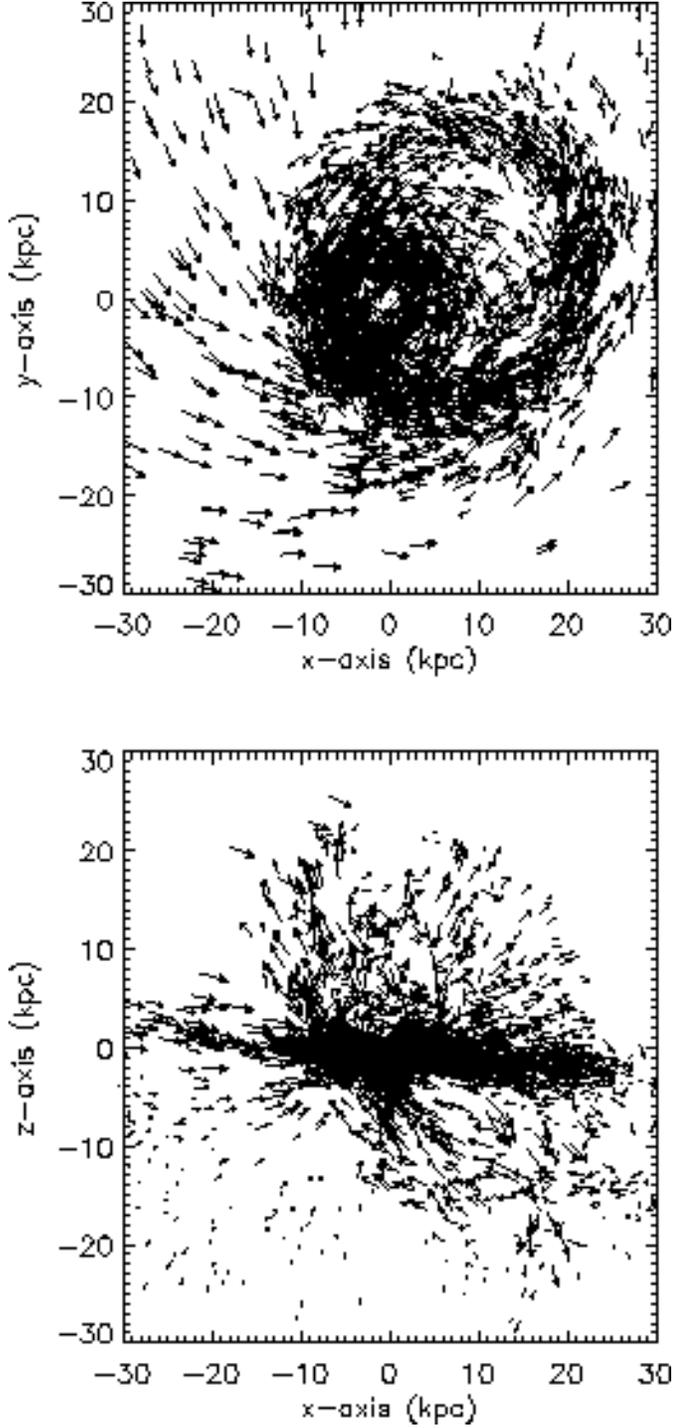}}
\caption{The gas cloud velocity field of the galaxy. The size of the
arrows are proportional to the cloud velocity. For clarity only
10\% of the arrows are plotted. Upper panel: Face-on
 view. Lower panel: Edge-on view.
} \label{fig:point_vel}
\end{figure}

The three-dimensional velocity field can be seen in Fig.~\ref{fig:point_vel}.
The galaxy rotates counter-clockwise. The gas which is located in the north
is accreting back to the galaxy. We observe a lot of counter-rotating clouds
in the expanding ring. In particular, there is a streamer which runs almost
perpendicular to the disk (Fig.~\ref{fig:point_vel} lower panel).
It comes from below and crosses the disk at its eastern edge.
We want to stress here again that there are a lot of clouds whose trajectories
will lead to collisions.

Since we have used a galaxy with an optical diameter of $D$=30 kpc
in our simulations (see Vollmer et al. 2000) whereas NGC~4522
has an optical diameter of $D_{25}$=18 kpc, we adapted the length
scale of our simulations to that of NGC~4522 (roughly a factor 2).
This three-dimensional cloud distribution and velocity field can now be
projected using the observed position angle and inclination.
We will directly compare the outcomes of this projection with the
observational results. The model images have almost the resolution
of the observations.

\subsection{The distribution of the ionized gas}

\begin{figure}
\resizebox{\hsize}{!}{\includegraphics{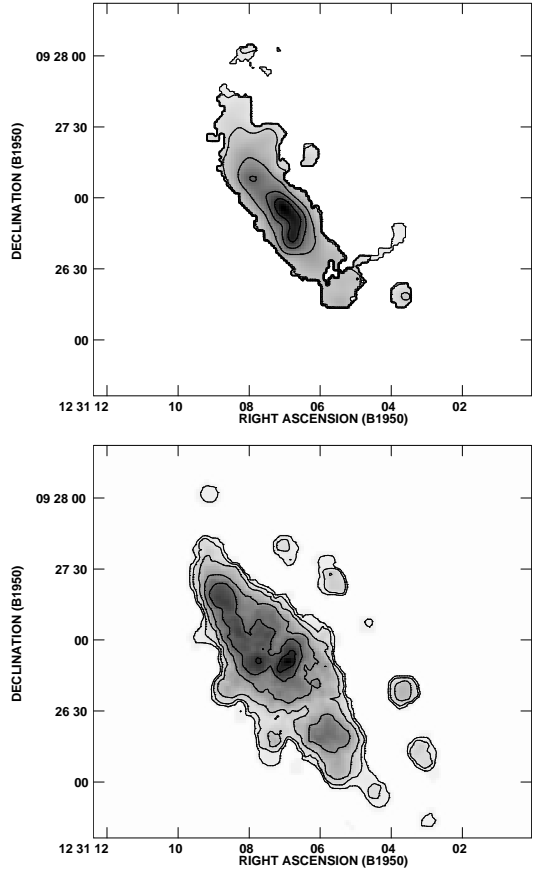}}
\caption{H$\alpha$ emission distribution. Upper panel:
observations. Lower panel: model.
} \label{fig:distributions}
\end{figure}

In order to model the H$\alpha$ distribution, we assume that star formation is
induced by the compression of the clouds during cloud--cloud collisions.
Consequently, we have to search for clouds with crossing trajectories in
order to detect gas streams which lead to frequent inelastic collisions.
We thus project the velocity vectors  (${\bf v_{1}}$, ${\bf v_{2}}$)
of each pair of clouds on their relative distance vector
${\bf r_{12}}$: $v_{1}^{\rm proj}={\bf v_{1}}\cdot
{\bf r_{12}}\ ;\ v_{2}^{\rm proj}={\bf v_{2}}\cdot {\bf r_{12}}\ .$
We apply a weighting factor to each cloud which is equal to
the number of possible collisions fulfilling the criterion:
$r_{12} \leq 1$ kpc, $v_{1}^{\rm proj}$ has the opposite sign of
$v_{2}^{\rm proj}$, and $R > 10$ kpc, where $R$ is the distance of the
cloud to the galaxy center. At distances $R \leq 10$ kpc, i.e. within
the disk, the mean free path of a cloud is smaller than outside the disk.
The criterion $r_{12} \leq 1$ kpc might thus not be valid at small distances.
Moreover, in the inner disk star formation is more likely due to
density waves within the stellar and gaseous disk. Since this aspect is not
included in the model we prefer to give the clouds at $R \leq 10$ kpc a
uniform weight. The model emission inside the galaxy disk is thus
density weighted, the emission outside the disk is collision weighted.
This treatment ensures in our view the simplest approach in order to
model the observed H$\alpha$ emission. Fig.~\ref{fig:distributions} shows
the observed together with the model of the H$\alpha$ emission distribution.

The observed H$\alpha$ emission distribution is already discussed in detail
in Kenney \& Koopmann (1999). We therefore just summarize the main features.
\begin{itemize}
\item
The H$\alpha$ disk is strongly truncated.
\item
10 \% of the total H$\alpha$ emission arises from a one-sided, extra planar
distribution.
\item
The extra planar ionized gas is organized into filaments.
\item
The extra planar H$\alpha$ emission has also a diffuse component.
\item
The region of highest H$\alpha$ emission in the core of the disk is more
extended to the south-west.
\item
The region of intermediate H$\alpha$ emission in the galaxy disk is more
extended to the north-east.
\end{itemize}

Almost all characteristics of the observed H$\alpha$ emission distribution
can also be found in the model distribution (Fig.~\ref{fig:distributions}
lower panel):
\begin{itemize}
\item
The disk is strongly truncated. This is due to the truncation of the
H{\sc i} gas disk at a radius $R \sim 7$ kpc.
\item
A fraction of several percent of the total emission can be found in the
north-west, far away from the disk. This emission is due to colliding
clouds which form stars ionizing their environment. This happens
predominantly in the expanding ring.
\item
Our model image does not have enough resolution to show filaments.
A careful inspection of the observed H$\alpha$ image of
Kenney \& Koopmann (1999)
shows that the filaments are mainly located near the north-eastern and
south-western end of the galaxy's disk. These are the places where
the expanding ring joins the disk. Moreover, the filamentary structure
can be due to magneto--hydrodynamic phenomena which are not included in our
code.
\item
We have not yet included the mechanism of star formation in the model.
We therefore can not make the difference between clumped and diffuse
emission directly. Nevertheless, since the expanding ring has a considerable
extension in the disk plane with a relatively low cloud density, we expect
that there is a part of UV emission which escapes out of the cloud
complex, where the stars were built. This environment
can be other clouds in the expanding ring or the tenuous material which
is distributed around the galaxy disk (see Fig.~\ref{fig:point_dist}).
The diffuse component of the observed H$\alpha$ emission might therefore
be the trace of this tenuous gas component.
\item
The maximum emission of the disk shows the same asymmetry as the
observed H$\alpha$ emission distribution.
\end{itemize}

\subsection{The velocity field of the ionized gas}

\begin{figure}
\resizebox{\hsize}{!}{\includegraphics{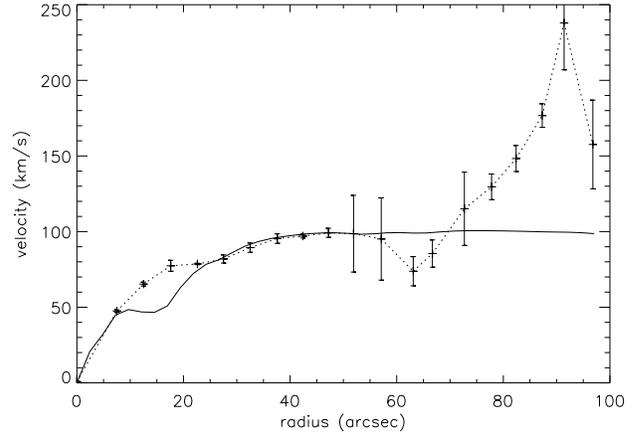}}
\caption{H$\alpha$ rotation curve averaged for both sides. The error bars correspond
to the difference between the approaching and the receding side.
Dotted line: observed rotation curve. Solid line: model rotation curve
(maximum disk model, R luminosity profile, M/L=5).
} \label{fig:mdm_model}
\end{figure}

We will first discuss the overall behavior of the observed H$\alpha$ rotation curve
averaged for both sides. For the comparison with our model, the approaching and
receding sides are separated. The  H$\alpha$ rotation curve can be seen in
Fig.~\ref{fig:mdm_model}.
It is made with a constant position angle and
a constant inclination angle. The error bars correspond to the difference between
the approaching and the receding side. The rotation curve
rises rapidly up to a distance of 20$''$ and more slowly for distances between
20$''$ and 50$''$. Between 50$''$ and 60$''$ the velocities of the approaching and
receding side differ considerably ($\sim$50\%). A drop of the rotation velocity
from 90~km\,s$^{-1}$ to 70~km\,s$^{-1}$ can be observed at 65$''$. The
rotation curve rises then linearly up to a distance of 90$''$. This rise is
due to the H$\alpha$ blobs located in the north west direction above the disk's
major axis.
The difference between both sides is quite small ($\sim$20~km\,s$^{-1}$)
for this outer part of the rotation curve.

The mass distribution  has been analyzed using a mass model and a code
written by Carignan (1985).
The solid line in Fig.~\ref{fig:mdm_model} corresponding to this model rotation curve
is based on the R luminosity profile (Koopmann \& Kenney 2000), an intrinsic ratio
Qo=0.15, a constant M/L=5, and a maximum disk assumption.
The mass surface density of the disk is derived from
the surface brightness profile at each radius. This explains the dip in
the model rotation curve at 15$''$ caused by a dip in the R luminosity profile, which
is due to the galaxy's internal structure, i.e. the bar.
The model curve reproduces accurately the H$\alpha$ rotation curve up to 60$''$.
Outside, it was impossible to reproduce the rising solid body rotation curve
using a maximum disk mass model or a best fit model with
any known shape of dark halo.  We therefore conclude
here that the H$\alpha$ emission blobs which are located in the north west must be
located beyond the disk plane and/or must be accelerated.

\begin{figure}
\resizebox{\hsize}{!}{\includegraphics{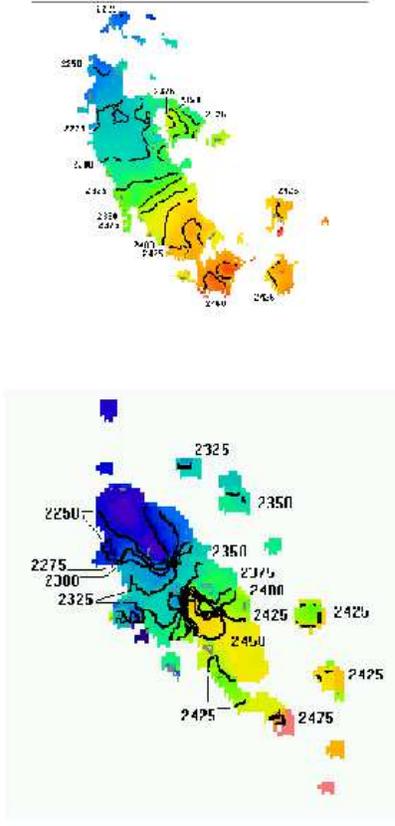}}
\caption{The H$\alpha$ velocity field. Upper panel:
observations. Lower panel: model.
The color code for the radial velocities is shown on top
of each panel.
} \label{fig:vfields}
\end{figure}

In the following paragraph the observed and the model rotation curves will be
directly compared.
Fig.~\ref{fig:vfields} shows the velocity field of the H$\alpha$ observations
together with that of the model for the approaching and receding side separately.
The overall structure of both velocity fields is in good agreement.
The difference in the inner disk is due to the initial model rotation curve,
which rises more rapidly than the observed one.

\begin{figure}
\resizebox{\hsize}{!}{\includegraphics{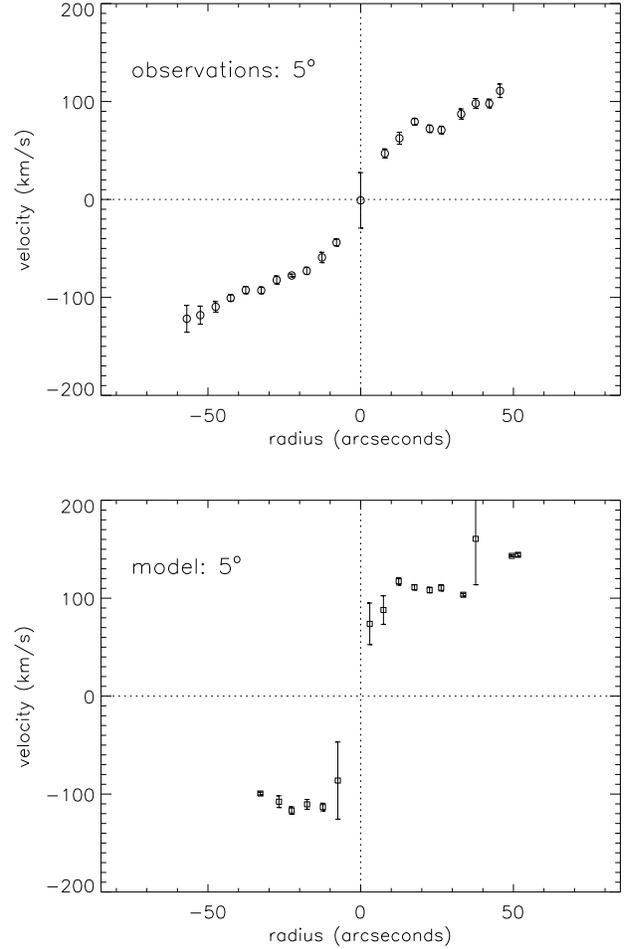}}
\caption{Derived rotation curves within a sector of 5$^{\rm o}$
around the major axis. Distances are positive to the south-west.
The error bars indicate the velocity dispersion in each ring.
} \label{fig:compare5}
\end{figure}

The derived rotation curves within a sector of 5$^{\rm o}$ around the major
axis is plotted in Fig.~\ref{fig:compare5}.
Our observations confirm the non-perturbed rising rotation
curve measured by Rubin et al. (1999).

The observed rotation curve continues to rise at both sides up to a distance
of $|R|\sim 50''$. The receding side shows a little bump at
$R \sim 20''$. The model rotation curve does not differ from the
initial model rotation curve for radii $R < 40''$, i.e. it stays constant.
The outer end at positive distances
begins to rise, whereas the outer end at the opposite side shows the inverse
trend.

As mentioned in the previous section, the initial model rotation curve is not
rising but flat. This changes obviously the velocity field of the inner
disk but not that for larger distances to the galaxy center. The main
result of the simulation at this point is that the rotation curve of the
inner disk is not altered by the ram pressure stripping event, i.e.
it stays flat.

\begin{figure}
\resizebox{\hsize}{!}{\includegraphics{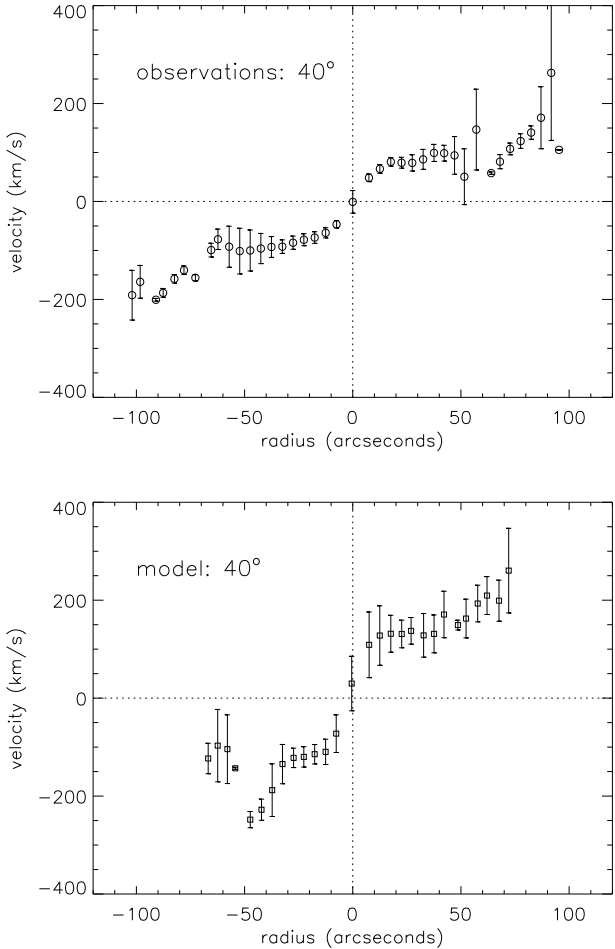}}
\caption{Derived rotation curves within a sector of 40$^{\rm o}$
around the major axis. Distances are positive to the south-west.
The error bars indicate the velocity dispersion in each ring.
} \label{fig:compare40}
\end{figure}

The most important test for the model velocity field is the velocity
structure of the emission at the north-eastern side of the galaxy disk.
These regions are included in the rotation curve derived within a sector
of 40$^{\rm o}$ around the major axis (Fig.~\ref{fig:compare40}).

The observed rotation curve is symmetric. It rises steeply up to a distance
of $|R| \sim 20''$ from the galaxy center. Then, the velocity increases more
slowly up to a distance of $|R|\sim 50''$. After a sudden drop it rises
linearly (solid body rotation) and reaches $\sim 200$ km\,s$^{-1}$ at a
distance of $|R| \sim 100''$.

The model rotation curve shows the same overall behavior. It rises
linearly at both sides and covers the same velocity range than the
observed rotation curve.
The main difference between the model and the observed rotation curve is
the asymmetry of the model rotation curve. While the receding side
has a very similar behavior as the observed rotation curve,
the linear rise of the rotation curve at the approaching side begins
at a smaller distance ($R \sim -30''$). For $R < -50''$ the rotation
curve drops abruptly. The observed rotation curve shows also a
decrease but at larger radius ($R \sim -90''$).

\subsection{Off-plane material and predicted H{\sc i} gas distribution and
velocity field}

\begin{figure}
\resizebox{\hsize}{!}{\includegraphics{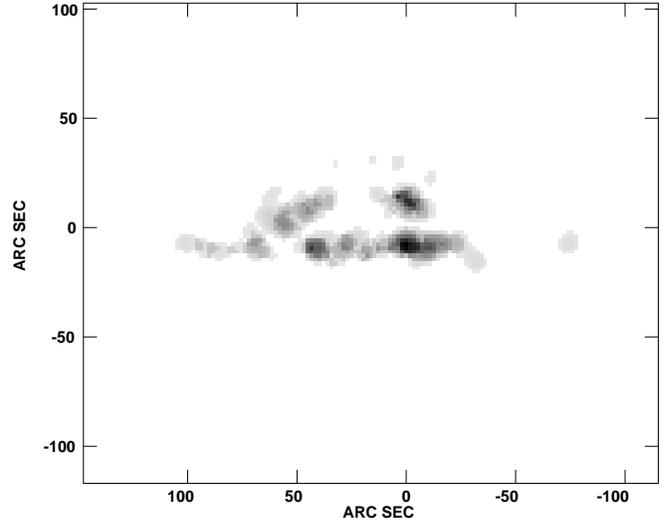}}
\caption{The model emission distribution seen edge-on. The clouds
located within the disk are removed from the model.
} \label{fig:edge-on}
\end{figure}

We can investigate if the material in the expanding ring is
located beyond the galaxy's disk plane as suggested by
Kenney \& Koopmann (1999). In order to study the emission
distribution in the $z$ direction we removed the disk particles from
the model and plotted the emission distribution for the galaxy plane
seen edge-on (Fig.~\ref{fig:edge-on}).
A substantial part of the gas clouds in the expanding ring located at
radii $R > 10$ kpc is thus well situated up to 2 kpc above the disk plane.

\begin{figure}
\resizebox{10cm}{!}{\includegraphics{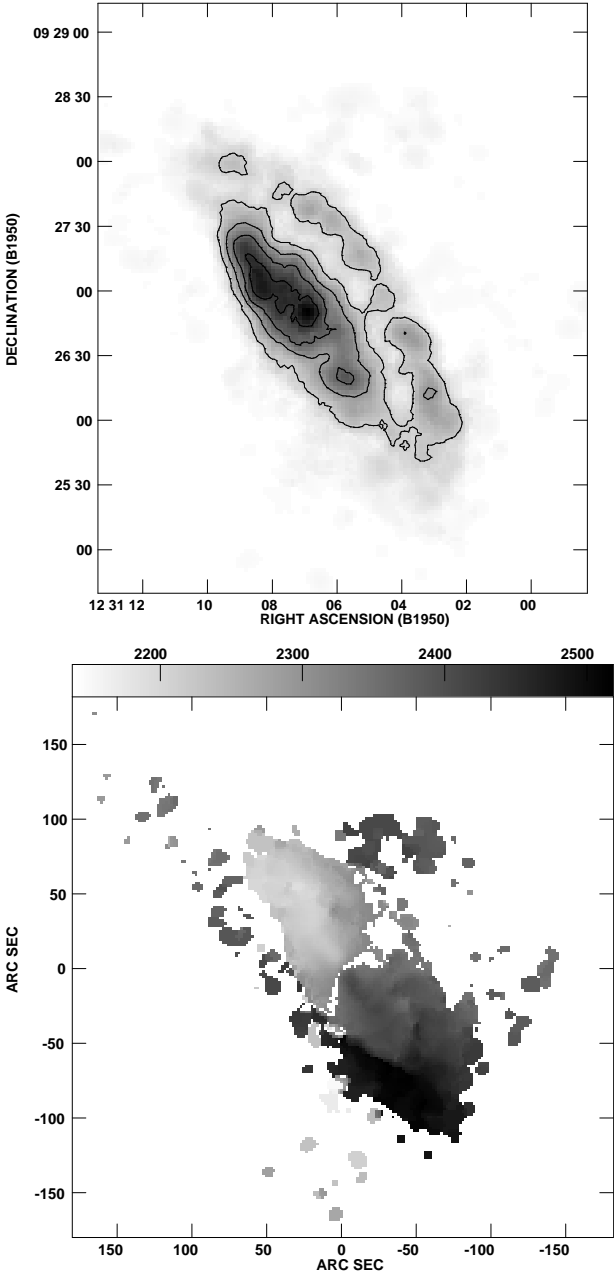}}
\caption{The H{\sc i} distribution and velocity field as it would
  be detected by very deep H{\sc i} 21 cm observations.
} \label{fig:HI_all}
\end{figure}

With the help
of the three-dimensional model cloud distribution and velocity field
we can attempt to predict the overall gas distribution and velocity field
as it would be detected by very deep H{\sc i} 21 cm observations
(Fig.~\ref{fig:HI_all}).
In doing so, we have to be very cautious. The major caveat for this prediction
is the unknown mass loss rate of the neutral gas clouds which are located
in the hot ICM due to evaporation. Since the stripping event has already
happened a long time ago, it is possible that a significant fraction of the
extra planar neutral gas clouds has already evaporated and is no longer
observable in H{\sc i}. Very deep H{\sc i} observations (VLA D array) will
therefore not only be a test for the chosen model but they will also give
some clues on the evaporation timescale of the stripped neutral gas clouds.

\section{Discussion and conclusions}

We have shown new H$\alpha$ Fabry-Perot interferometer observations of the
Virgo cluster spiral galaxy NGC~4522. The velocity field can be traced up
to radius of $\sim 60''$. We compare the results of these observations
with a numerical model which includes the effects of ram pressure acting
on the galaxy's gas content.

We derived rotation curves within a sector of 5$^{\rm o}$
and 40$^{\rm o}$ around the major axis. Thus, we have been
able to compare the emission distribution, the velocity field, and the
rotation curves of the observations with those of the model.

For the inner disk of the galaxy, the main result is that the ram pressure
stripping event which happened 6.5\,10$^{8}$ yr ago did not alter the
initial rotation curve of the galaxy. For the outer disk, the stripped
material which has not been accelerated above the escape velocity is
re-accreting on the galaxy. The majority of this material forms an expanding,
asymmetric ring structure. A small fraction of the stripped gas
is falling back onto the galaxy within a counter-rotating arm.
Inelastic collisions between the gas clouds located in the ring
on the one hand and between clouds of the counter-rotating arm and
clouds located in the ring on the other hand happen frequently.
This leads to an enhanced star formation activity within the ring structure.
Since we have traced the total number of collisions during the simulation,
we can estimate the star formation rate due to cloud--cloud collisions:
during the last $\Delta t=$5\,10$^{7}$ yr the number of collisions is
$n_{\rm coll}$=31. With a cloud mean mass
$\overline{m}_{\rm cl}$=8\,10$^{5}$ M$_{\odot}$ and assuming that
10 \% of the gas is turned into stars this gives
\begin{equation}
SFR = 0.1\,n_{\rm coll}\,2\,\overline{m}_{\rm cl}/\Delta t \simeq 0.1
\ {\rm M}_{\odot}\,{\rm yr}^{-1}
\end{equation}
This is in excellent agreement with the measured star formation rate
of $SFR = 0.11\ {\rm M}_{\odot}\,{\rm yr}^{-1}$ (Kenney \& Koopmann 1999).
The newly formed massive and
hot stars ionize their surroundings giving rise to the H$\alpha$ emission.
These detached emission blobs lead generally to a rise of the rotation curve,
and sometimes to a decreasing rotation curve. This corresponds
to the behavior of the observed velocity field.

It is important to mention that the numerical simulation does not pretend
to reproduce the H$\alpha$ distribution and velocity field in each detail.
The aim of the comparison is to find the same general characteristics which allow
us to discriminate whether the model describes the observations well. In our case,
the comparison shows the following similarities:
\begin{itemize}
\item
The H$\alpha$ disk is strongly truncated.
\item
A considerable amount  of the total H$\alpha$ emission arises from a
one-sided, extra planar distribution.
\item
The region of highest H$\alpha$ emission in the core of the disk is more
extended to the south-west.
\item
The region of intermediate H$\alpha$ emission in the galaxy disk is more
extended to the north-east.
\item
The rotation curves of the inner disk are those of an undisturbed galaxy.
\item
The rotation curves of the outer disk shows a linear rise (solid body
rotation).
\item
The detached emission blobs lead generally to a rise of the rotation curve,
and sometimes to a decreasing rotation curve.
\end{itemize}
The dissimilar aspects are by far less numerous:
\begin{itemize}
\item
The observed H$\alpha$ emission off-plane comes mainly from three distinct
regions, whereas the model emission off-plane covers 180$^{\rm o}$.
\item
The model rotation curve is not symmetric.
\item
The model emission distribution does not show a filamentary structure.
\item
The model emission distribution does not show a diffuse component.
\end{itemize}
The last two points are due to numerics (discrete model, missing resolution,
no magnetic fields, no star formation mechanism).
The asymmetry of the model rotation
curve is most likely due to the initial conditions of the simulation.
We have run several simulations with different initial conditions, but
since the parameter space for these initial conditions is very large,
we were not able to find a model snapshot which fits our observations
more tidily.

We therefore conclude that the model describes very well the overall
aspects of the H$\alpha$ observations.

This leads to the final conclusion that the galaxy's closest passage to
the cluster center is $\sim$6.5\,10$^{8}$ yr ago and the galaxy is coming
out of the cluster core. Its positive radial velocity with respect to
the cluster mean velocity places it behind the cluster center (M87).

\begin{acknowledgements}
The authors wish to thank Jacques Boulesteix for his help on the data
acquisition system and data reduction software for GHASP.
BV is supported by a TMR Programme of the European Community
(Marie Curie Research Training Grant).
\end{acknowledgements}

\end{document}